\documentclass[runningheads]{llncs}
\usepackage[T1]{fontenc}
\usepackage{graphicx}
\usepackage{hyperref}
\usepackage[backend=biber,style=ieee]{biblatex}

\def\BibTeX{{\rm B\kern-.05em{\sc i\kern-.025em b}\kern-.08em
    T\kern-.1667em\lower.7ex\hbox{E}\kern-.125emX}}
\begin{document}
\title{The Lifecycle Workbench - A Configurable Framework for Digitized Product Maintenance Services}
\titlerunning{The Lifecycle Workbench}
%
\author{Dominique Briechle \inst{1}\orcidID{0009-0000-2610-3399} \and
Mohammed Fahad Ali \inst{1}\orcidID{0009-0000-5048-2173} \and
Marit Briechle-Mathiszig \inst{1}\orcidID{0009-0003-8571-4212} \and
Tobias Geger \inst{1}\orcidID{0009-0004-4469-534X} \and
Robert Werner \inst{1}\orcidID{0009-0007-2516-0536} \and
Andreas Rausch \inst{1}\orcidID{0000-0002-6850-6409}}

\authorrunning{D. Briechle et al.}
\institute{Clausthal Unitversity of Technology, 38678 Clausthal-Zellerfeld, Germany 
\email{\{dominique.fabio.briechle, mohammed.fahad.ali, marit.elke.anke.mathiszig, thomas.tobias.marcello.geger, robert.werner, andreas.rausch\}@tu-clausthal.de}}

%
\maketitle              
\begin{abstract}
The global production of electric goods is at an all-time high, causing negative environmental and health impacts as well as a continuing depletion of natural resources. Considering the worsening global climate change, a transition of current industrial processes is necessary to tackle the above-mentioned factors. To address this urgent issue, socio-economic systems like the Circular Economy (CE) provide options to reallocate the use of resources and products on a global scale. Especially in terms of product lifecycle-prolonging, this system provides suitable approaches to alter the current modes of product handling by society and industry alike, based on the condition of the products. Although the importance and benefits of sustainable services enabling these options are widely known, users tend to shy away from using them. One of the reasons is the missing reliability in terms of the knowledge of the costs associated with a particular service. This uncertainty in expected pricing can, therefore, lower the willingness of potential clients. However, not only clients struggle with the boundary conditions of such services. On the part of the potential providers of services, the monetary risk is often caused by the incapability to detect the condition of a product in advance. This can result on the provider side in a severe economic loss if this possibility is not covered by the service price or through the mass of items, which could allow equalization of serval service operations.  
To address these weak points in current service execution, the authors propose the \textit{Lifecycle Workbench (LCW)}-ecosystem, which features digital representations to enhance the reliability of service pricing as well as the assessment of the condition of items, assemblies, and parts in the Circular Economy domain.

\end{abstract}

\section{Introduction}\label{sec:Introduction}

Global production of electric and electronic goods is rising every year and is expected to exceed 9000 million products by 2029, which is 1500 million more than back in 2018 \cite{statistaconsumerelectronics}. This continuously rising production causes severe environmental and health impacts for both nature and humans alike. For example, pollutants generated during the production of plastic compounds \cite{noh2022environmental}, the depletion of natural resources as well as the amount of energy required for the manufacturing process of new products \cite{2023Accelerating} represent threats to the environment. In addition to this unsustainable production behavior, the global amount of electric and electronic waste (WEEE) is rising with every year, exceeding the 60 million metric tons mark in 2022 \cite{balde2024global}. 

In contrast to this development, where products are deposited after reaching their primary end of life (EoL) based on criteria like damages, technical obsolescence or no longer desired design features, the Circular Economy as an economic model aims to prolong the use of products and resources alike. For products, it leverages certain options which are directed to enhance the longevity of the same including reuse, repair, remanufacturing, and repurposing \cite{kirchherr2017conceptualizing}.

However, wide-spanning socio-economic models like the Circular Economy face several challenges when it comes to their implementation, especially in the area of reliability and risk connected to the determination of the cost of a certain service \cite{grafstrom2021breaking}. This becomes apparent by looking at the two main stakeholder types in such a system: the user and/or owner of a product, called in the following \textit{ProductAdministrator}, and the provider, called in the following \textit{ServiceProvider}, who offers services based on the options proposed by the Circular Economy model to prolong the product lifecycle. 

One reason why \textit{ProductAdministrators} do not have repaired their devices is the lack of reliability of the cost determination of such services, caused by the incapability of \textit{ServiceProviders} to assess the products before they have received them. Before sending an item to a \textit{ServiceProvider}, the to-be-expected costs are, therefore, often not certain or are based on fixed prices, which de-incentivizes potential service users because of either high or unknown costs they are not willing to pay or an absence of economical feasibility. This lack of uncertainty can be extended as well to the to-be-expected repair time, which depends on several factors like free capacity on the \textit{ServiceProvider}-side, availability of spare parts and level of knowledge of the \textit{ServiceProvider}. These factors are, however, not directly assessable by the \textit{ProductAdministrator}, which increases the de-motivation to utilize such services. 

On the \textit{ServiceProvider}-side, the main barrier for offering services is the high amount of risk caused by the uncertainty of actual cost on his side \cite{oh2016impact}. This is especially the case considering CE-services, where the information density regarding the targeted product is often insufficient. Since the assessment of a product is often hardly possible before a \textit{ProductAdministrator} ships it to the \textit{Serviceprovider}, the nature of the product's malfunction is often unknown to him.  

A potential solution to increase the reliability for both stakeholders could be the deployment of digital ecosystems, which make use of digital representations as stakeholder proxies. These representations are capable, with the help of supporting tools that assess the condition of an incoming product, to generate precise service offers in the scope of the configuration of a \textit{ServiceProvider} while allowing the \textit{ProductAdministrator} to accept those offers based on his defined configuration. 

In the scope of the paper, the authors are presenting a possible approach for such a digital ecosystem concept: the Lifecycle Workbench. To address the highlighted shortcomings of modern approaches and point out the interactions in the proposed \textit{LCW}-ecosystem, the authors are focusing on a framework proposal for a digital service system model and its process conduction flow, shown in a scenario in \autoref{sec:lcw-tooling-examples}. Thereby, the following two research questions are addressed: 

\begin{itemize}
 \item \textit{\textbf{RQ1}: How can digitization ensure the reliability of service offers in the product maintenance service domain?}
 \item \textit{\textbf{RQ2}: How can the requirements of different stakeholder types in product maintenance services be determined?} 
 
\end{itemize}

The paper is structured in the following way:
\autoref{sec:scenario} describes a motivational scenario to point out the hurdles of nowadays service systems, whereas \autoref{sec:conception} presents the framework of the \textit{LCW} and the anticipated ecosystem. In \autoref{sec:lcw-tooling-examples}, the initially described scenario is mapped to the proposed framework solution. The section is followed by \autoref{sec:discussion and future work}, in which the research questions, the systems limitations and future research directions of the proposed \textit{LCW}-framework are discussed. Finally, the paper summarizes with the conclusion in \autoref{sec:conclusion}.

\section{Motivation \& Scenario} \label{sec:scenario}
As already highlighted in \autoref{sec:Introduction}, one of the identified main barriers of Circular Economy-related, product lifecycle-prolonging services are the reliability in terms of cost estimation and compliance with time constraints. Although the awareness regarding the importance of CE services as a tool to achieve a more sustainable society is already on a high level, this circumstance hinders and demotivates potential \textit{ProductAdministrator} from using such services. \textit{ServiceProviders} are, however, often not capable of naming an accurate price estimation for a service. This creates a dilemma for most CE-\textit{ServiceProviders} nowadays, that offer their services based on one of two business models. 

The first model presupposes the \textit{ProductAdministrator} to send the regarded product, which can be either an \textit{Item, Assembly, or Part}, to the \textit{ServiceProvider}. The division into these three distinct subgroups concludes from the way physical products are normally built up and helps us understand scenarios, where not a product as a whole is the focal point of the service. The \textit{ServiceProvider} assesses the same before making an offer for the reinstation of the functionality. 

\begin{figure}[tbh]
  \includegraphics[width=\linewidth]{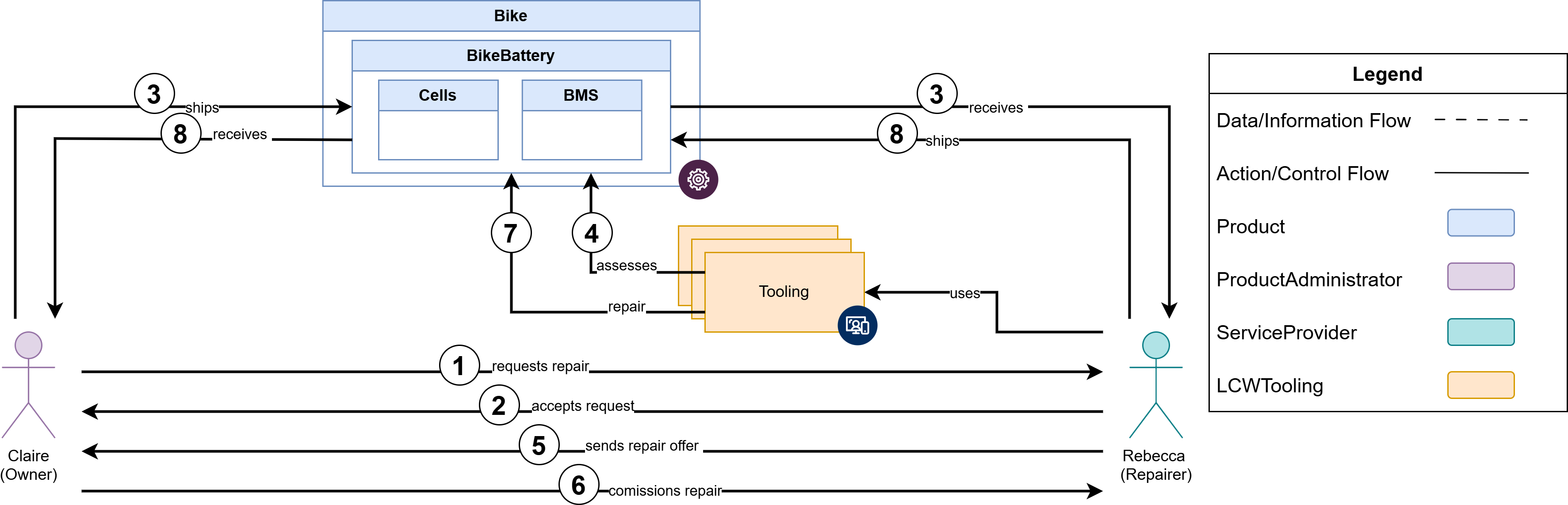}
  \caption{Initial Bike battery repair scenario}
  \label{fig: Defect battery without LCW} 
\end{figure}

The second model gets by without this necessity by offering a fixed value for a particular service. The \textit{ServiceProvider} is therefore trying to cover his expenses by distributing the costs among the incoming dysfunctional products. The logic behind this approach is to balance the cost by naming an average price for the conduction of a particular service, which features a price margin. This, however, represents a potential risk since too many severely damaged or not serviceable products will decrease the revenue of the \textit{ServiceProvider}, if he still wants to offer the service to the \textit{ProductAdministrator} for an acceptable price.

To highlight the identified issues, the authors designed a scenario, acting as an initial problem description, which visualizes the necessity of the development of the anticipated \textit{LCW}-framework for sustainable service design. For the scenario, the authors selected a product that is featured as a core component in E-bikes. 

E-bikes currently experience an all-time high in terms of urban and suburban mobility, especially in larger urban areas, for example, Dublin \cite{hosseini2024bike}. The scenario, shown in Fig. \ref{fig: Defect battery without LCW}, describes the interaction of two stakeholders, our E-bike \textit{ProductAdministrator} Claire, as well as Rebecca, the \textit{ServiceProvider}. Claire recognizes that her E-bike battery is depleted and is no longer capable of storing enough energy for her daily commute to her job. To change this, she is in search of somebody who can repair the battery. Luckily for Claire, after a while, she finds on the Internet a suitable \textit{ServiceProvider}, Rebecca, who is able to repair her battery. She is, therefore, sending Rebecca a request for the repair of the battery \textbf{(1)}. After a day, Rebecca has time to look up her emails and reviews Claire's request. She accepts the repair job \textbf{(2)} but is for now not able to assess the potential cost of the operation, a circumstance she draws Claire's attention to. Claire sends in the absence of options the battery nevertheless to Rebecca in the hope that the repair won't be too expensive \textbf{(3)}.
Rebecca receives the battery and after an additional day, where she had to take care of other clients of hers, she starts with the measurements of the battery cells in order to determine the damage \textbf{(4)}. After the assessment of the damage, Rebecca has now determined the root of the malfunction and calculates the repair cost. She is writing a repair offer and sends it to Claire \textbf{(5)} via E-Mail. Claire reviews the offer and is shocked at how costly the repair of her battery is. Unfortunately, she is in dire need of her battery and accepts contritely the offer after a day of reflection time \textbf{(6)}. Unfortunately, Rebecca has to wait until the spare parts are delivered to conduct the repair. After five more days, the spare parts are finally delivered to her and she can start with the repairing of the battery \textbf{(7)}. After a total of 8 days, the battery's functionality has been reestablished, it is shipped back to Claire \textbf{(8)}. Claire receives her battery after 4 more days and can use her bicycle again. Although Claire is happy, that she can now use her bike again, she is as well frustrated because of the way the repair process was handled. 

This small scenario points out the two major obstacles of such service systems, which are addressed by the framework analogue to the aforementioned research questions: 
\begin{itemize}
    \item The precise estimation of service cost for unknown or not fully known situations, as shown in the scenario, is an ubiquitous and long-known obstacle for the acceptance of lifecycle-prolonging services and product-service systems in general, as mentioned in \cite{erkoyuncu2009uncertainty} and \cite{erkoyuncu2011understanding}. The dissatisfaction of Claire, our \textit{ProductAdministrator}, could have been avoided by (1.) directly sending her a reliable repair offer for her battery and (2.) providing her with comparable alternative offers from different \textit{ServiceProviders}. Although, in contrast to individual, case-based pricing of service cost, a general fixed price could've prevented this scenario as well, it would've potentially caused dissatisfaction because of an unfair price in the perception of Claire, our \textit{ProductAdministrator} or to a loss-making procedure for Rebecca, our \textit{ServiceProvider}.
    
    \item In addition to service costs, the timely constraints play a major role in the acceptance of service offers. As highlighted in the scenario in \ref{sec:scenario}, these services can, however, require a certain duration to be fulfilled, spanning from days to weeks \cite{GUSSERFACHBACH2023137763}, because of the elaborate process of handling, assessing, and repairing products.
    Thereby, this will directly influence the cost of those services and the acceptance by the \textit{ProductAdministrator}. Because of the high cost and the expected downtime in product use, potential users can be discouraged to rely on such services. Whereas this circumstance is already tackled in certain industries, for example in automotive services \cite{genzlinger2020servitization}, with the opportunity to book a replacement car for the service time, it is not an integral part of nowadays maintenance service design in general. However, to increase the attractiveness of maintenance services for a diverse palette of products, the safeguarding of timely constraints, and by that the lowering of costs for conducting the service, is essential for a broad application of Circular Economy-based maintenance services like the anticipated \textit{LCW}.  
    
\end{itemize}

\section{\textit{LCW}-framework Conception} \label{sec:conception}
\begin{figure}[tbh]
  \includegraphics[width=\linewidth]{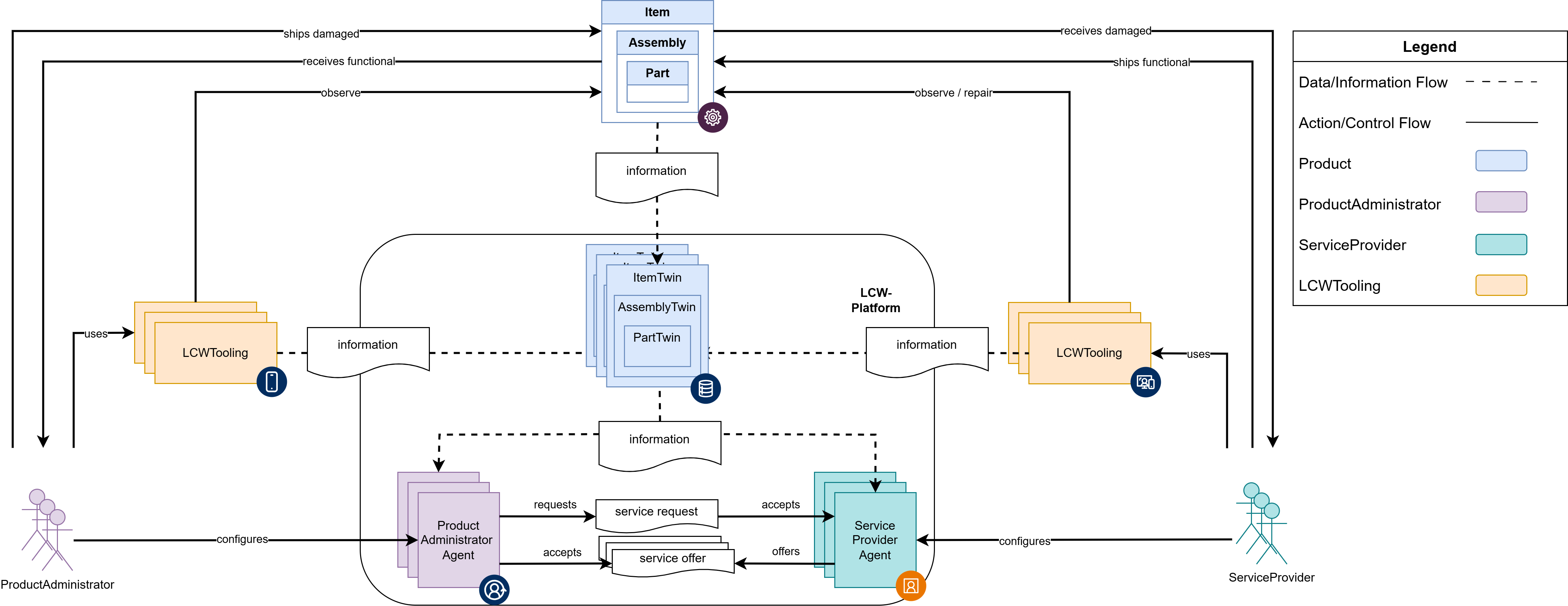}
  \caption{\textit{LCW}-framework describing the interaction between the different stakeholders and system components}
  \label{fig: LCW-Model} 
\end{figure}

The scenario in \autoref{sec:scenario} has pointed out the two major barriers which prevents both the stakeholders from using or offering CE services for the prolonging of product lifecycles. The proposed \textit{LCW}-framework, shown in Fig. \ref{fig: LCW-Model}, has, therefore, to ensure both the generation of precise price estimations on the part of the \textit{ProductAdministrator} as well as the mitigation of financial risk on the part of the \textit{ServiceProvider}. Further, the \textit{LCW}-framework needs to ensure that the identified temporal delays, caused by the stakeholder's interaction with one another as well as the selected business models, are addressed. In general, as one of the major protagonists of the ecosystem, the \textit{ProductAdministrator} represents various types of product users, owners, and managers, for example, private owners, fleet managers, and product distributors. The \textit{ProductAdministrator} has full control over one or more products, divided in this context into \textit{Item, Assembly and Part}, and is the final deciding entity when it comes to the terms of triggering the service process and accepting the service offer from the \textit{ServiceProvider}. Depending on the nature of the product, the \textit{ProductAdministrator} is further the systems' entity responsible for controlling the information flow and distribution towards different entities, for example, the \textit{ItemTwin} in the \textit{LCW}. 

The second kind of stakeholder, the \textit{ServiceProvider}, represents different types of service providers that carry out the lifecycle-prolonging operations according to their business model and ensure the reinstatement of the functionality of the \textit{Item, Assembly, or Part}. These service operations can reach from the usual service interval of an item for a checkup to the restoration of the same by building it up from scratch. However, as will be shown in \autoref{sec:lcw-tooling-examples}, the functionality for a \textit{ProductAdministrator} of an \textit{Item, Assembly, or Part} as such can also be reinstated through measures of exchange. Thereby, the \textit{ProductAdministrator} will receive a spare product instead of his previously owned, that will offer the same functionality. The proposed framework is designed to accompany different business models for CE-Services targeted at the prolonging of product lifecycles.

Both stakeholders have at their disposal tools that enable the assessment of the product, the \textit{LCWTooling}. For the \textit{ServiceProvider}, the \textit{LCWTooling} can take on many shapes and features a manifold of characteristics, spinning from simple tools for repairing to automated repair facilities and cloud-backed Artificial Intelligence (AI)-based tools for the evaluation of a large variety of different products. On the \textit{ServiceProvider}-side, the tooling, therefore, includes both tools for the assessment and tools for the functionality reinstatement, hence the repair. For the \textit{ProductAdministrator}, the \textit{LCWTooling} will be most likely an application, which is executable on mobile phones and home computers. In contrast to the tools of the \textit{ServiceProvider} the purpose is solely the assessment of the product's condition. Both \textit{LCWToolings} are, in addition, used to distribute its collected information to the Digital Twin of the corresponding \textit{Item, Assembly or Part}. 

\indent The central element of the proposed \textit{LCW}-ecosystem is the \textit{LCW}-platform. The \textit{LCW-Platform} provides a frame for the integration of the digital representations of both the \textit{ProductAdministrator} and the \textit{ServiceProvider} as well as the Digital Twin of the \textit{Item, Assembly or Part}. For both the stakeholders these representations act as digital proxy, outfitted to take over the management of the service conduction. In contrast, the Digital Twin of the product acts as the core information hub for the received transmission streams. These information are transmitted to the Twin from the \textit{LCWTooling}, used by both stakeholders, and, if the product supports a form of connectivity, from the \textit{Item, Assembly or Part} directly. 

\indent In contrast to the Twins of the \textit{Items, Assemblies, or Parts}, the digital representations of both the \textit{ProductAdministrator} and the \textit{ServiceProvider} are active agents. This means that they can actively manipulate other entities in the ecosystem based on their configuration. Stakeholders can, therefore, use these agents and define the rules, upon they will initiate certain actions. Both the agents are granted access to the Twins of the \textit{Item, Assembly, or Part} to be able to fulfill their and derive the according actions. As depicted in Fig. \ref{fig: LCW-Model}, the \textit{ProductAdministratorAgent} has two fundamental actions, it can request and accept a service offer within the platform. The \textit{ProductAdministrator}, therefore, configures the boundary conditions on whom the \textit{ProductAdministratorAgent} can act to find an according \textit{ServiceProvider}. These boundary conditions can be, for example, financial or timely constraints. The \textit{ServiceProviderAgent} on the other part crawls the active service requests and matches those to its own configuration, generated by the \textit{ServiceProvider}. With the active search of suitable service requests, the \textit{ServiceProviderAgent} is granted full access to the information about the product, that is stored in the corresponding Twin and is linked to the service request. The \textit{ServiceProviderAgent} features as well conditions that lay down the rules of interaction for themselves. These can be, for example, the price of a service and the time it will take for fulfillment or a specification of the kind of product a service is offered for. If a service request matches the own conditions, the \textit{ServiceProviderAgent} will act by sending a service offer to the \textit{ProductAdministratorAgent}. The \textit{ProductAdministratorAgent} will collect the incoming offers and will then decide based on his constraints, which one will be selected for the conduction of the service.

\section{\textit{LCW}-framework Scenario} \label{sec:lcw-tooling-examples}

In this section, the defined framework presented in the previous chapter is now applied to the scenario introduced in \autoref{sec:scenario} to show its applicability. The scenario features here an exchange part business model. As already mentioned in \autoref{sec:conception}, this means that the \textit{ServiceProvider} does not send the original received product back to the \textit{ProductAdministrator}. Rather, a product of the same type is used in this process to reinstate the functionality on the part of the \textit{ProductAdministrator}. The original received product is repaired after the finalization of the exchange process by the \textit{ServiceProvider} and is stored by the same in order to be available for future service requests. 

\begin{figure}[tbh]
  \includegraphics[width=\linewidth]{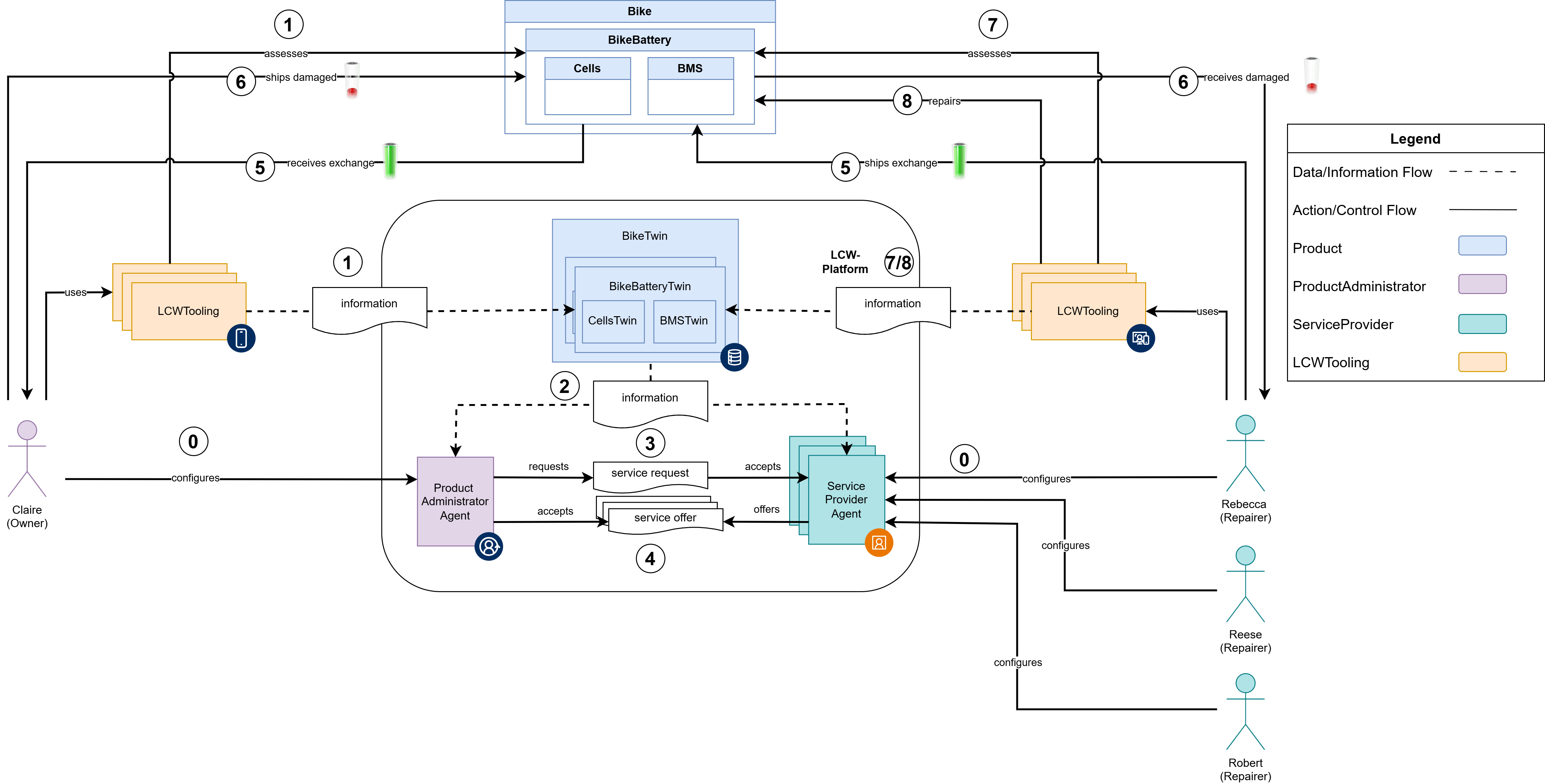}
  \caption{Scenario description applying the \textit{LCW}-framework to maintain a battery repair service for clients}
  \label{fig: Defect battery with LCW} 
\end{figure}

Again, the initial situation for our scenario is the defective \textit{BikeBattery}. However, in contrast to the scenario shown in Fig. \ref{fig: Defect battery without LCW}, we now apply the framework presented in \autoref{sec:conception} to design the \textit{LCW}-application as shown in Fig. \ref{fig: Defect battery with LCW}.
In this scenario, our \textit{ProductAdministrator} is again Claire, who owns the defective product, the \textit{Bikebattery}. Our \textit{ServiceProvider} Rebecca is as well again part of the scenario, and offers a service for the reinstation of the functionality. However, in contrast to the introduced scenario in \autoref{sec:scenario}, Rebecca is no longer the only \textit{ServiceProvider}, that offers such a service. Together with her, Robert, and Reese offer as well services that reinstate the functionality of the product.
Both the stakeholder groups have their respective \textit{LCWTooling} at their disposal, which allows the assessment and, in the case of the  \textit{ServiceProvider}, the repair of the product.

Before the actual process is started, the different stakeholders configure their respective agents \textbf{(0)}. Claire on her part configures her \textit{ProductAdministratorAgent} with the boundaries of the later generated service request. She, therefore, defines parameters like the maximum cost she is willing to pay for the reinstation of the \textit{BikeBattery's} functionality as well as the maximum duration she is willing to wait for the service fulfillment. For her \textit{BikeBattery} she is willing to pay 400 € and requests the functionality reinstation in a duration of 6 days. The three \textit{ServiceProvider} featured in the scenario, Robert, Reese and Rebecca, all offer services for the reinstation of batteries. Therefore, they have configured their \textit{ServiceProviderAgents} to match the conditions of the services they offer. To simplify the scenario, the three have configured their agents as well with just the two conditional parameters cost and time. The three \textit{ServiceProviderAgents} are featuring the following configuration:
\begin{itemize}
    \item Rebecca's Agent: Duration and cost for functionality reinstation: 14 days \& 350 €
    \item Robert's Agent: Duration and cost for functionality reinstation: 5 days \& 450 €
    \item Reese's Agent: Duration and cost for functionality reinstation: 4 days \& 400 €
\end{itemize}

Now that the agents are configured to act on the stakeholder's behalf, the process itself can be conducted.
The process is initially triggered by Claire through the assessment and damage detection of the \textit{BikeBattery}, using her \textit{LCWTooling} device. In this scenario, Claire uses an application on her mobile phone as \textit{LCWTooling}, which allows the optical recording of the product's features.
With the conclusion of the assessment of the \textit{BikeBattery}, the \textit{BikeBatteryTwin} is updated with the recorded information \textbf{(1)}. According to recorded data, the type and manufacturer of the battery are determined as well as damages on the outside of the battery casing. In this scenario, the \textit{LCWTooling} detects a damage on the main connection plug of the \textit{BikeBattery}.
At first, the \textit{ProductAdministratorAgent} receives the information stored in the \textit{BikeBatteryTwin}, which are used to help to formulate the service request \textbf{(2)}. 

The \textit{ProductAdministratorAgent} is now generating a service request, which is based on its own configuration as well as the information stored in the \textit{BikeBatteryTwin} \textbf{(3)}. Now the actively searching \textit{ServiceProviderAgents} have access to the offer request information and can decide, based on the available information, if they want to send a service offer to the \textit{ProductAdministratorAgent} \textbf{(4)}. Since all the \textit{ServiceProviderAgents} are configured in a way that the corresponding service is directed at the specific \textit{BikeBattery} model, all of them are sending a service offer with the previously introduced parameters. Since Reese's service offer is based on an exchange model where a refurbished \textit{BikeBattery} is exchanged for a malfunctioning one, her agent is able to offer the reinstation of functionality within the shortest service duration while still being within Claire's defined price margin. Therefore, the \textit{ProductAdministratorAgent} accepts the offer of Reese's \textit{ServiceProviderAgent}.

Reese is acknowledged by her \textit{ProductAdministratorAgent} of the acceptance of Claire's agent and sends her a refurbished \textit{BikeBattery} of the same model \textbf{(5)}. Claire is now able to use her E-bike again and is happy with the easy and uncomplicated way this process is handled. By receiving the new \textit{BikeBattery}, her \textit{LCWTooling} is automatically connected to the corresponding \textit{BikeBatteryTwin}. Claire is sending her damaged \textit{BikeBattery} back to Reese and does not have to worry about the disposal of the same \textbf{(6)}.

However, for Reese, the process continues after receiving the \textit{BikeBattery} \textbf{(7)} with the assessment of the same in detail by using the \textit{LCWTooling}. The tool assesses the condition, in this case, measuring the capacity of cells and the functionality of the BMS, etc. and transmits the information to the \textit{BikeBatteryTwin}. Following the assessment, Reese uses again her \textit{LCWTooling} to conduct the repair of the \textit{BikeBattery}, namely the exchange of certain cells and the BMS as well as the exchange of the damaged plug by disassembly and dissolving the connections with automated tools, like an automated Soldering Tool, in \textbf{(8)}. After finishing the repair, Reese stores the \textit{BikeBattery} in her warehouse for future service requests. 

\section{Discussion \& Future Work} \label{sec:discussion and future work}

The depicted scenario and the proposed framework describe, how digital ecosystems can support not only the dissemination and optimization of already exciting services but also the conception of new business models enabled by digital technologies. This is especially the case for mechanics like the presented exchange model in \ref{sec:lcw-tooling-examples}. Regarding the two initially raised research questions, the scenario and framework are able to provide some insight for the discussion.


\begin{itemize}
    \item {\textbf{RQ1}: In terms of the question regarding the reliability of such service offers, the two core issues highlighted in the motivation are addressed by the framework through digital representations: the assurance of the reliability of an offer for the \textit{ProductAdministrator}, as well as the risk mitigation on part of the \textit{ServiceProvider} due to an increase in information provided by the \textit{LCWTooling}. Further, for future clients the reliability paired with the additional benefits, like a potential decrease in service cost as well as a reduction in service time, can provide a more appealing alternative in contrast to classical service offers in the sustainability domain. This especially becomes apparent when comparing the proposed \textit{LCW}-ecosystem with approaches, where neither the cost nor the time for the functionality reinstatement are known at the time the service is commissioned. 
    
    However, the foundation for an accurate offer generation is the \textit{LCWTooling}, which can consist of different types of technologies, that support the gathering of information. The significance of the \textit{LCW}-ecosystem is, therefore, highly dependent on the assessment capabilities of the different tools summarized in the \textit{LCWTooling}. The selection of tools is, however, scenario-dependent and can vary widely for different products and stakeholder configurations. Further, as depicted in the framework in \autoref{sec:conception}, certain products feature transmission technologies, which allow the independent update of the corresponding Digital Twin by the product itself, which can enhance the accuracy of the automated offer generation. The elucidation of the \textit{LCWTooling} is held in the scope of the paper rather abstract and is only described in the context of the scenario. This is due to the high degree of possible scenarios as well as the variation in \textit{ServiceProviders}, that can be involved in reinstating the functionality of a product. In the scenarios, rather basic \textit{LCWTooling} applications, like measuring devices or simple repair units, were introduced for the sake of comprehensibility. However, in the context of the research for sustainable and digital ecosystems more complex systems for Circular Economy based services are investigated as well. To further increase the processing speed of such services, the authors proposed an automated disassembly system, which could support the pre-processing of products in order to accelerate operations like repairing, remanufacturing or repurposing \cite{Briechle2024}. The \textit{LCW}-framework could offer, therefore, an integration of several smart tools to support the \textit{ServiceProvider} even further by automating the maintenance process on a larger scale. However, for the application of the ecosystem on an application level, the \textit{LCWTooling} is highly dependent on both the stakeholders and the products alike. The proposed framework, therefore, could help define these requirements and support the development of technologies necessary to assess the incoming products.
    
    Regarding the assessment of the framework itself, the described scenario allows a theoretical comprehensibility of the proposed, system-wise mechanics of how service offers could be generated. To finally determine the impact and potential effect on cost saving and productivity with regards to the service offer reliability, the framework needs to be implemented in a real context or a simulation environment under defined economic key performance indicators. An assessment can, therefore, only be given for the theoretic-technical application, whereas the economic effects still have to be investigated. Future research in this direction should, therefore, focus on how to specify the interaction mechanics between the stakeholders to enable the in-depth evaluation of the framework in that regard.} 
    
    
    \item {\textbf{RQ2}: Whereas the first research question targets the evaluation of the \textit{LCW}-ecosystems' overall applicability, the second one deals with the methodological application of the developed framework. Frameworks in general are proven to be capable of greatly supporting the development process of digital systems \cite{morisio1999framework}\cite{valerio1997domain}. In the context of the anticipated \textit{LCW}-ecosystem, the framework provides the systematic foundation, that allows the further development of the system's components. This is especially the case with regard to the different types of Agents and Digital Twins, which have to be adjusted to the needs of the given use cases as well as the \textit{LCWTooling}. The framework is offering at this stage an insight, into how the stakeholders in the system are interacting with one another. This knowledge can be used to support the design of the different entities on an application level, by defining the scope of functionality necessary for the specific component. 
    
    Especially for both the Twins and agents the application of the framework to highlight a certain business model, as shown in \ref{sec:lcw-tooling-examples}, enables insights into the definition of the required parameter necessary for the configuration. in the context of the scenario, a rather simple configuration was selected for comprehensibility's sake. When imagining larger use cases, these representations will become more complex and can feature various parameters, extending the presented ones for cost time and product type. 

    The same applies to the Digital Twins, which are logically increasing in complexity analog with more complex and fine-particle products. This becomes apparent when comparing the product used in our scenario, the \textit{BikeBattery} to its parent product, the E-bike. This increasing complexity will require the inclusion of more components of the parent product into its digital representation and will, therefore, make the modeling of the same more challenging.
    
    The assurance of the stakeholder requirements is, therefore, a complex topic which can be supported by the knowledge provided by the proposed \textit{LCW}-framework. Together with the contained information about the interaction mechanics in the framework and the constraints of certain business models, the definition of the requirements for the different stakeholders can be supported. The emphasize for future research should lay, therefore, on further breaking down the requirements of the stakeholders and investigate several real-world applications to define the scope of the same. In addition, the specification of the configuration of the digital representations in that regard should be further explored.}
   
\end{itemize}

Beyond the above mentioned directions for the future research, the framework provides several points of contact for adjunct services in the periphery of the \textit{LCW}-ecosystem.

Especially when considering subsequent services, like product logistics in this context, it becomes apparent, that the framework in its current state relies on common logistic channels, like the national post services or multinational shipping services. However, considering the two factors time and sustainability, other means of transportation could be integrated into the model to elevate the efficiency over multiple domains of the process and reduce the potential repair time by further cutting down the non-productive time. 
For example, smart logistic concepts for the \textit{LCW}-ecosystem could be investigated to raise on one hand the overall simplicity of the process for the \textit{ProductAdministrators}, while on the other enhancing a more sustainable logistic chain by incorporating concepts like inter-modal and multi-modal logistics. Especially when such concepts are tied together with a multi-provider ecosystem, capable of distributing rides, similar to a concept presented by Ali et al. in 2024 \cite{ali2024party} for the delivery of goods, this could further leverage sustainable services. With the help of the integration of a smart, distributed, and adaptable logistics concept, maintenance service systems could be thereby fleshed out in more ways, opening even new lines of business for potential \textit{ServiceProviders}.

\indent Another point of contact for subsequent research is the optimization of lifecycle decisions across the system including all stakeholders. In its current state, the system is designed as a framework for multiple \textit{ServiceProviders}, which are generating offers for a pool of potential clients, the \textit{ProductAdministrators}. The system enables the interaction between those stakeholders but does not offer any decision-making capabilities, which could support the \textit{ProductAdministrators} in its overall lifecycle decision. Moreover, it is targeted to support the \textit{ServiceProviders} and their digital representation \textit{ServiceProviderAgent} to derive a conclusion about the potential degree of functionality reinstatement they could provide under two restrictions: cost and time. However, in a more holistic approach to a Circular Economy, with its options to Reuse, Repair, Remanufacture, etc., the system lacks the opportunity to optimize the utilization of products, and by this, to distribute the resources to the most suitable \textit{ServiceProvider}, ecological and economically wise.
The \textit{LCW}-framework could, therefore, be developed in the future into a generalized lifecycle optimization platform. Assessment and optimization processes could be adapted to generate specific treatment decisions for a product, thereby enhancing the systems' functionality and potential contribution to a Circular Economy. This would advance the sustainability of the overall system by large since the lifecycle decision would no longer be dependent on the decision of the \textit{ProductAdministrators} himself. However, it would also decrease the control by the actors of the system, which could deter potential users on both sides.

\indent 

\section{Conclusion}\label{sec:conclusion}
The paper presents an approach for designing a product maintenance service ecosystem with the support of the \textit{LCW}-framework, which helps define the configurations and interactions of the different stakeholders and digital entities within \textit{LCW}-platform. An initial scenario was presented to motivate the research questions raised in \autoref{sec:Introduction}. The presented framework was defined and the main stakeholders and interactions within the ecosystem were described. In addition, an application of the framework was described in \autoref{sec:lcw-tooling-examples} to show the theoretical functionality of the same. However, although the framework provides a possible foundation for such systems, it opens up several possible connection points for subsequent research as discussed in \autoref{sec:discussion and future work}. Especially in preparation for a real-world application of the framework, the further research regarding the \textit{LCWTooling} and the different digital representations is crucial. 

\section{Acknowledgment}
This work has been developed in the project “Life\_TWIN” (Research Grant Number 03EI5014A) and is funded by the German Federal Ministry for Economic Affairs and Climate Action (BMWK). The following partners were involved in this project: 
Clausthal University of Technology, Robert Bosch GmbH, Hellmann Process Management GmbH \& Co. KG and Bernhard Olbrich Elektroinstallationen-Industrieanlagen GmbH.

\printbibliography

@article{GUSSERFACHBACH2023137763,
title = {Repair service convenience in a circular economy: The perspective of customers and repair companies},
journal = {Journal of Cleaner Production},
volume = {415},
pages = {137763},
year = {2023},
issn = {0959-6526},
author = {Ines Güsser-Fachbach and Gernot Lechner and Tomás B. Ramos and Marc Reimann}
}

@inproceedings{ali2024party,
  title={Party Without a Cake? Onto an Inter-modal HitchHike Logistics Platform for Passengers and Products Transportation},
  author={Ali, Mohammed Fahad and Briechle, Dominique and Briechle-Mathiszig, Marit and Geger, Tobias and Rausch, Andreas},
  booktitle={European Conference on Software Architecture},
  pages={100--114},
  year={2024},
  organization={Springer}
}

@article{2023Accelerating,
  title={Accelerating the circular economy in Europe - State and outlook 2024},
  author={zu Castell-Rudenhausen, Malin and Wahlstr{\"o}m, Margareta and Nelen, Dirk and Dams, Yoko and Paleari, Susanna and Zoboli, Roberto and Wilts, Henning and Bakas, Ioannis},
  year={2023},
  publisher={European Environment Agency (EAA)}
}

@article{kirchherr2017conceptualizing,
  title={Conceptualizing the circular economy: An analysis of 114 definitions},
  author={Kirchherr, Julian and Reike, Denise and Hekkert, Marko},
  journal={Resources, conservation and recycling},
  volume={127},
  pages={221--232},
  year={2017},
  publisher={Elsevier}
}

@article{hosseini2024bike,
  title={E-bike to the future: Scalability, emission-saving, and eco-efficiency assessment of shared electric mobility hubs},
  author={Hosseini, Keyvan and Choudhari, Tushar Pramod and Stefaniec, Agnieszka and O’Mahony, Margaret and Caulfield, Brian},
  journal={Transportation Research Part D: Transport and Environment},
  volume={133},
  pages={104275},
  year={2024},
  publisher={Elsevier}
}

@misc{statistaconsumerelectronics,
    title ={Consumer Electronics - Worldwide, , n.d.. [Online]},
    howpublished = {https://www.statista.com /outlook/cmo/consumer-electronics/worldwide},
    note = {Accessed: 2025-04-30}
    
}

@misc{balde2024global,
  title={Global e-waste monitor 2024},
  author={Bald{\'e}, Cornelis P and Kuehr, Ruediger and Yamamoto, Tales and McDonald, Rosie and D’Angelo, Elena and Althaf, Shahana and Bel, Garam and Deubzer, Otmar and Fernandez-Cubillo, Elena and Forti, Vanessa and others},
  year={2024},
  publisher={International Telecommunication Union (ITU) and United Nations Institute for~…}
}

@article{grafstrom2021breaking,
  title={Breaking circular economy barriers},
  author={Grafstr{\"o}m, Jonas and Aasma, Siri},
  journal={Journal of cleaner production},
  volume={292},
  pages={126002},
  year={2021},
  publisher={Elsevier}
}

@article{Briechle2024,
    author = {Briechle, Dominique and Rausch, Andreas},
    title = {{You’ve Got a Plan? A Domain Modelling Approach for Collaborative Product Disassembly Planning with PDDL}},
    journal = {ADVCOMP 2024, The Eighteenth International Conference on Advanced Engineering Computing and Applications in Sciences},
    year = {2024}
}

@article{noh2022environmental,
  title={Environmental and human health risks of plastic composites can be reduced by optimizing manufacturing conditions},
  author={Noh, Yoorae and Odimayomi, Tolulope and Sendesi, Seyedeh Mahboobeh Teimouri and Youngblood, Jeffrey P and Whelton, Andrew J},
  journal={Journal of Cleaner Production},
  volume={356},
  pages={131803},
  year={2022},
  publisher={Elsevier}
}

@article{genzlinger2020servitization,
  title={Servitization in the automotive industry: How car manufacturers become mobility service providers},
  author={Genzlinger, Felix and Zejnilovic, Leid and Bustinza, Oscar F},
  journal={Strategic Change},
  volume={29},
  number={2},
  pages={215--226},
  year={2020},
  publisher={Wiley Online Library}
}

@inproceedings{erkoyuncu2009uncertainty,
  title={Uncertainty challenges in service cost estimation for product-service systems in the aerospace and defence industries},
  author={Erkoyuncu, JA and Roy, R and Shehab, E and Wardle, P},
  booktitle={Proceedings of the 1st CIRP IPS2 Conference, Cranfield},
  pages={200--206},
  year={2009}
}

@article{erkoyuncu2011understanding,
  title={Understanding service uncertainties in industrial product--service system cost estimation},
  author={Erkoyuncu, John Ahmet and Roy, Rajkumar and Shehab, Essam and Cheruvu, Kalyan},
  journal={The International Journal of Advanced Manufacturing Technology},
  volume={52},
  pages={1223--1238},
  year={2011},
  publisher={Springer}
}

@article{oh2016impact,
  title={Impact of cost uncertainty on pricing decisions under risk aversion},
  author={Oh, Sechan and Rhodes, James and Strong, Ray},
  journal={European Journal of Operational Research},
  volume={253},
  number={1},
  pages={144--153},
  year={2016},
  publisher={Elsevier}
}

@inproceedings{morisio1999framework,
  title={Framework based software development: investigating the learning effect},
  author={Morisio, Maurizio and Romano, Daniele and Moiso, Corrado},
  booktitle={Proceedings Sixth International Software Metrics Symposium (Cat. No. PR00403)},
  pages={260--268},
  year={1999},
  organization={IEEE}
}

@article{valerio1997domain,
  title={Domain analysis and framework-based software development},
  author={Valerio, Andrea and Succi, Giancarlo and Fenaroli, Massimo},
  journal={ACM SIGAPP Applied Computing Review},
  volume={5},
  number={2},
  pages={4--15},
  year={1997},
  publisher={ACM New York, NY, USA}
}

%
%
%

\end{document}